\documentclass[journal,twocolumn,10pt]{IEEEtran}
\usepackage{setspace} 

\usepackage[T1]{fontenc}
\usepackage[T1]{fontenc}
\usepackage{amsmath,amssymb,amsfonts,mathrsfs,bm}
\usepackage{amsthm}
\usepackage{cite}
\usepackage{array}
\usepackage{enumerate}
\usepackage{graphicx}
\usepackage{psfrag}
\usepackage{url}
\usepackage{color}
\usepackage{algorithm,algorithmic}
\usepackage{multirow}
\usepackage[hidelinks]{hyperref}
\usepackage{breqn}
\newtheorem{assumption}{Assumption}

\newtheorem{theorem}{Theorem}
\newtheorem{lemma}{Lemma}

\newtheorem{lemma_A}{Lemma}[section]

\usepackage[normalem]{ulem}

\newcommand{\hide}[1]{}

\newcommand{\hhide}[1]{}
\newcommand{\cP}{\mathcal{P}}
\newcommand{\tn}[1]{^{(#1)}}

\makeatletter
\newcommand\dm@star[1]{\begin{dmath*}#1\end{dmath*}}
\newcommand\dm@nostar[2]{\begin{dmath}\label{#2}#1\end{dmath}}
\newcommand\dm[2][]{
  \ifthenelse{\equal{#1}{}}{
    \dm@star{#2}}{
    \dm@nostar{#2}{#1}}}
\makeatother

\begin{document}

\title{Robust Decentralized Detection and Social Learning in Tandem Networks}

\author{Jack~Ho,~\IEEEmembership{Student Member,~IEEE}, Wee~Peng~Tay,~\IEEEmembership{Senior Member,~IEEE}, Tony~Q.~S.~Quek,~\IEEEmembership{Senior Member,~IEEE}, and Edwin~K.~P.~Chong,~\IEEEmembership{Fellow,~IEEE}%
\thanks{Part of this paper was presented at the IEEE International Conference on Acoustics, Speech and Signal Processing, May 2013. This research is supported in part by the Singapore Ministry of Education Academic Research Fund Tier 2 grants MOE2013-T2-2-006 and MOE2014-T2-1-028. J.~Ho and W.P.~Tay are with the School of Electrical and Electronic Engineering, Nanyang Technological University, Singapore. E-mail: \texttt{JHO008@e.ntu.edu.sg, wptay@ntu.edu.sg}. T.Q.S.~Quek is with the Information Systems Technology and Design Pillar,  Singapore University of Technology and Design, Singapore. E-mail: \texttt{tonyquek@sutd.edu.sg}. Edwin K.P.~Chong is with the Department of Electrical and Computer Engineering, Colorado State University, USA. E-mail: \texttt{edwin.chong@colostate.edu}.}
}



\maketitle \thispagestyle{empty}


\begin{abstract}
We study a tandem of agents who make decisions about an underlying binary hypothesis, where the distribution of the agent observations under each hypothesis comes from an uncertainty class. We investigate both decentralized detection rules, where agents collaborate to minimize the error probability of the final agent, and social learning rules, where each agent minimizes its own local minimax error probability. We then extend our results to the infinite tandem network, and derive necessary and sufficient conditions on the uncertainty classes for the minimax error probability to converge to zero when agents know their positions in the tandem. On the other hand, when agents do not know their positions in the network, we study the cases where agents collaborate to minimize the asymptotic minimax error probability, and where agents seek to minimize their worst-case minimax error probability (over all possible positions in the tandem). We show that asymptotic learning of the true hypothesis is no longer possible in these cases, and derive characterizations for the minimax error performance.
\end{abstract}

\section{Introduction}
In this paper, we formulate and study the robust social learning problem in a tandem network. A tandem network consists of agents connected in a serial fashion, where each agent receives an opinion about a binary hypothesis from a previous agent, and makes a decision about a binary hypothesis based on the previous agent's opinion and its own observation. Despite the simple structure of the tandem network, studying it can lead to insights about more complicated network structures such as those in social networks or Internet of Things (IoT) networks. The tandem network approximates a single information flow in a network, and it and its variants have been widely studied in \cite{VisThoTum:88,TanPatKle:91b,Cov:69,HelCov:70,Kop:75,PapAth:92,Swa:93,TayTsiWin:08,AceDahLob:08,DraOzdTsi:13}. 

In our model, each agent's decision is based on a local error criterion, which it selfishly tries to optimize. This behavior is present in social networks, where users are mainly concerned with spreading only locally accurate information. In this paper, we call this \textit{social learning} \cite{Tay:J15,AceDahLob:08,SmiSor:00,Ban:12,BanFud:04,BikEta:92,ZhaPezMor:12,ZhaChoPez:13,DraOzdTsi:13}. One such application of social learning is in the case of participatory sensing, where inference about a phenomenon of interest is made through the help of agents in the network \cite{Lee2010, Demirbas2010, Tay:J15}. For example, this can occur when users send a picture of litter in a park to a social sensing platform \cite{Wang2013, Wang2} or report congested road conditions. \cite{Inrix}.

On the other hand, if the agents' decision rules are designed to minimize the error criterion of the last agent in the network, or the asymptotic error probability in the case of an infinite network, this is known as \textit{decentralized detection} \cite{Tsi:93,TayTsiWin:08,Var:97}. One major application of decentralized detection is in sensor networks with a fusion center \cite{ChaVar:86,VisVar:97,LinCheVar:05}. If the fusion center is able to relay information to the other agents, it will be able to select a set of globally optimal decision rules for every agent. However, many practical networks, such as social networks, do not have a fusion center. Furthermore, even for networks with fusion centers, the fusion center may not be able to easily communicate with the other agents. This is true of the participatory sensing examples above.

In the above examples, it is assumed that each agent knows the distribution of its private observation, and that of its predecessor, as well as its position in the network. However, in a real-life network, this is generally not the case. In this paper, we investigate what happens when one or both of these assumptions do not hold.

\subsection{Related Work}

Binary hypothesis testing in a tandem network model is studied in \cite{Cov:69, Kop:75}, which shows that learning the true hypothesis asymptotically is possible with unbounded likelihood ratios, and not possible with bounded likelihood ratios when agents transmit only 1-bit messages. 
Decentralized detection policies for tandem networks are also considered in \cite{PapAth:92}, and conditions for the error probability approaching zero as the number of agents grows are derived. This is a network where each agent after the first receives exactly one decision from its predecessor. The authors also study a sub-optimal scheme where each sensor ``selfishly'' tries to minimize its own error, as opposed to the error of the root agent. The reference \cite{TayTsiWin:08} shows that the rate of error decay is at most sub-exponential. 

Feedforward networks, in which an agent obtains information from a subset of previous agents not necessarily just the immediate predecessor, have been studied in \cite{AceDahLob:08,ZhaChoPez:13,DraOzdTsi:13}. In \cite{DraOzdTsi:13}, agents are able to access the decisions of their $K$ most recent predecessors. It is demonstrated that almost sure learning is impossible for any value of $K$, and learning in probability is possible for $K\geq 2$. A new model where forward looking agents try to maximize the discounted sum of the probability of a right decision is also considered. The reference \cite{AceDahLob:08} studies the decentralized detection problem in a game theoretic setting, and examines the effect of obtaining information from different sets of previous agents on the rate of error decay. The reference \cite{ZhaChoPez:13b} examines the asymptotic error rate of feedforward topologies under two types of broadcast errors, namely erasure and random flipping.



All the above works assume that agent's observations are drawn from \emph{known} distributions under each hypothesis. This assumption may not hold in practical networks like IoT networks, in which sensors' observation distributions may change over time, or in social networks, in which agents' observations may be affected by the agents' mood at a particular time. The robust detection framework was first proposed by \cite{Hub:65} for a single agent to model the case where the observation distributions are not known exactly. A survey of results in this area can be found in \cite{KasPoo:85}. The underlying probability distributions governing the agent observations are assumed to belong to different uncertainty classes under different hypotheses, and it is shown that under a minimax error criterion, the optimal decision rule for the agent is a likelihood ratio test based on the pair of least favorable distributions (LFDs). Subsequently, the work \cite{VeeBasPoo:94} investigates robust detection in a finite parallel configuration, with and without a fusion center. The problem of robust sequential detection is studied in \cite{GerCha:90}. Robust social learning however has not been addressed in these works. In addition, robust detection and learning have not been studied for the tandem network.

\subsection{Our Contributions}

In this paper, we consider robust binary hypothesis detection and social learning in a tandem network in which the observation models of agents under each hypothesis are uncertainty classes of probability distributions. Our main goals are to obtain the optimal agent policies under a minimax error criterion, under both decentralized detection and social learning frameworks, and characterize the asymptotic minimax error probabilities under various scenarios. Our main contributions are the following:
\begin{enumerate}
\item For the tandem network, we show that the solutions to the robust decentralized detection and social learning problems are equivalent to the respective solutions of the corresponding classical hypothesis testing problem where all the private observations are distributed according to the LFDs (Theorem \ref{theorem:LFD_tandem}). Our proof can be extended to general tree topologies, which generalizes a result in \cite{VeeBasPoo:94}, where a parallel topology is considered. 

\item We show that when the uncertainty classes for all agent observations are the same, and agents know their positions in the tandem, asymptotically learning the true hypothesis under both decentralized detection and social learning frameworks is not possible if the contamination of both uncertainty classes are non-zero, even when the log likelihood ratio of the nominal distributions is unbounded (Theorem \ref{theorem_asymptotic} and Theorem \ref{theorem:asymptotic_social}). This is in contrast to the case where the contamination of the uncertainty classes are zero \cite{PapAth:92,TayTsiWin:08}, in which case asymptotic learning happens if the log likelihood ratio is unbounded.

\item When agents know their positions in the network, we show that asymptotically learning the true hypothesis under social learning is achievable if and only if the log likelihood ratio of the nominal distributions is unbounded, and there are two subsequences of agents, one corresponding to each hypothesis, such that the contamination of the uncertainty class under that hypothesis converges to zero (Theorem \ref{theorem:varying}).

\item When agents do not know their positions in the tandem, we show that it is not possible to asymptotically learn the true hypothesis. We investigate the cases where agents collaborate to minimize the asymptotic minimax error probability, and where agents seek to minimize their worst-case minimax error probability (over all possible positions in the tandem), and characterize the minimax error performance in these approaches (Theorems \ref{theorem:unknown_decentralized} and \ref{theorem:unknown_social}). 
\end{enumerate}

In this paper, we consider only robust decentralized detection and social learning in tandem networks, which are very simple in structure. Social networks and IoT networks are much more complex in practice. Therefore, our results are limited, and can only be applied \emph{heuristically} to more practical networks. Our analysis forms the foundation for studying more complex networks like trees and general loopy graphs, and provides insights into designing optimal decision rules for such networks. For example, although using likelihood ratio tests based on LFDs at each agent is not known to be optimal for loopy graphs, we expect this to produce reasonable results in practice. Addressing the performance of social learning in more complex networks is part of future research.

The rest of this paper is organized as follows. In Section \ref{section:formulation}, we introduce the robust decentralized detection and social learning problem in a tandem network. In Section \ref{section:robust}, we provide a characterization for the agents' optimal decision rules in both decentralized detection and social learning in a tandem network. We then derive necessary and sufficient conditions for asymptotically learning the true hypothesis under various simplifications in Section \ref{section:asymptotic}. We also study the case where agents do not know their positions in the tandem in this section. In Section \ref{section:numeric}, we illustrate some of our results using a numerical example. Lastly, we conclude in Section \ref{section:conclusion}.

\section{Problem Formulation}\label{section:formulation}
\begin{figure}[!t]
\centering
\includegraphics[width=0.45\textwidth]{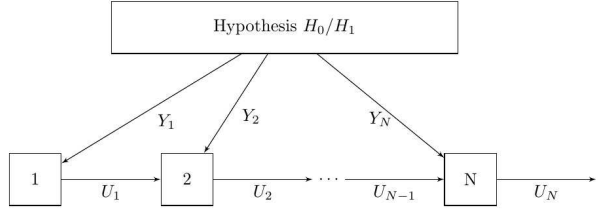}
\caption{Hypothesis testing in a tandem network.\label{fig:tandem}}
\end{figure}

We consider a tandem network consisting of $N$ agents, with agent 1 being the first agent and $N$ being the last (see Figure \ref{fig:tandem}). Consider a binary hypothesis testing problem in which the true hypothesis $H$ is $H_i$ with prior probability $\pi_i \in (0,1)$, for $i=0,1$. Conditioned on $H=H_i$, each agent $k$ in the network makes an observation $Y_k$, defined on a common measurable space $(\mathcal{Y},\mathcal{A})$, and with distribution $P_{i,k}$ belonging to an uncertainty class
\begin{dmath*}
\cP_{i,k} =\left\{{Q \mid  Q =(1-\epsilon_{i,k})P_i^* + \epsilon_{i,k} R}, {R \in \mathcal{R}}\right\},
\end{dmath*}
where $\mathcal{R}$ is the set of all probability measures on $(\mathcal{Y},\mathcal{A})$, $P_i^* \in \mathcal{R}$ is the nominal probability distribution, and $\epsilon_{i,k} \in [0,1)$ is a positive constant that is sufficiently small so that $\cP_{0,k}$ and $\cP_{1,k}$ are disjoint. We assume that all distributions in $\cP_{0,k}$ and $\cP_{1,k}$ are absolutely continuous with respect to one another, and the distribution $P_{j,k}$ from which the observation $Y_k$ is drawn from is unknown. Furthermore, we assume that conditioned on the true hypothesis, the observations of each agent are independent from one another. The parameter $\epsilon_{i,k}$ is also known as the \emph{contamination} for the uncertainty class $\cP_{i,k}$. When $\epsilon_{i,k}=0$, we recover the classical Bayesian hypothesis testing problem. While agents can have different contamination values $\epsilon_{0,k}$ and $\epsilon_{1,k}$, we assume that the nominal distributions $P_0^*$ and $P_1^*$ are identical for every agent.

For $k=1,\ldots,N$, each agent $k$ makes a decision $U_k = \phi_k(Y_k, U_{k-1}) \in \{0,1\}$ about the hypothesis $H$, where $\phi_k$ is an agent decision rule whose decision $i$ corresponds to deciding in favor of $H_i$, and $U_0 \equiv 0$. For $i=0,1$, let $P_i^{(N)}=P_{i,1}\times P_{i,2}\times ... \times P_{i,N}$. Similarly, define $\mathcal{P}_i^{(N)}=\mathcal{P}_{i,1}\times \mathcal{P}_{i,2}\times ... \times \mathcal{P}_{i,N}$. In the decentralized detection problem, our aim is to find a sequence of decision rules $\phi^{(N)}=(\phi_1,\phi_2,...,\phi_N)$ to minimize the maximum probability of error given by 
\begin{dmath}
P_{N}^{\text{DD}}(\phi^{(N)}) =  \sup_{(P_0\tn{N},P_1\tn{N})\in\mathcal{P}_0\tn{N}\times\mathcal{P}_1\tn{N}} P_{e,N}(\phi\tn{N}, P_0\tn{N}, P_1\tn{N}),\label{detection_prob}
\end{dmath}
where
\begin{dmath}\label{error_prob}
P_{e,N}(\phi\tn{N}, P_0\tn{N}, P_1\tn{N}) = \pi_0 P_{F,N}(\phi^{(N)},P_0^{(N)})
+\pi_1P_{M,N}(\phi^{(N)},P_1^{(N)}). 
\end{dmath}
In \eqref{error_prob}, $P_{F,N}$ and $P_{M,N}$ are the false alarm and missed detection probabilities of agent $N$ respectively, given the decision rules $\phi\tn{N}$ and the  agents' observation distributions $P_i\tn{N}$, $i=0,1$.

In the social learning problem, the first agent chooses its decision rule $\phi_1$ to minimize 
\begin{dmath*}
\sup_{(P_{0,1},P_{1,1})\in\mathcal{P}_{0,1}\times\mathcal{P}_{1,1}} P_{e,1}(\phi_1, P_{0,1},P_{1,1}).
\end{dmath*}
Each other agent $k$, given the decision rules $\phi\tn{k-1}$ of the previous agents $1,\ldots,k-1$, is able to derive the false alarm and miss detection probabilities of agent $k-1$. It then seeks to find $\phi_k$ to minimize
\begin{dmath}
P_{k}^{\text{SL}}(\phi_k \mid  \phi\tn{k-1}) =  \sup_{(P_0\tn{k},P_1\tn{k})\in\mathcal{P}_0\tn{k}\times\mathcal{P}_1\tn{k}} P_{e,k}(\phi\tn{k}, P_0\tn{k}, P_1\tn{k}). \label{social_learning}
\end{dmath}
In contrast to the decentralized detection problem in \eqref{detection_prob}, each agent \emph{myopically} seeks to minimize its local maximum probability of error.

For each agent $k$, let $p_{i,k}$ be the density (with respect to some common measure) of $P_{i,k}$, and $p_i^*$ be the density of $P_i^*$, for $i=0,1$. The least favorable distributions (LFDs) for two given uncertainty classes $\cP_{0,k}$ and $\cP_{1,k}$ is defined in \cite{Hub:65} to be the pair of distributions $(Q_{0,k}, Q_{1,k})$ with densities $(q_{0,k}, q_{1,k})$ such that
\begin{dmath*}
q_{0,k}(y) =\begin{cases}
(1-\epsilon_{0,k})p_0^*(y) &\textnormal{   for    } p_1^*(y)/p_0^*(y)<c'' \\  (1-\epsilon_{0,k})p_1^*(y)/c'' &\textnormal{       for      } p_1^*(y)/p_0^*(y) \geq c'' 
\end{cases}
\end{dmath*}
\begin{dmath*}
q_{1,k}(y) =\begin{cases} (1-\epsilon_{1,k})p_1^*(y) &\textnormal{   for    } p_1^*(y)/p_0^*(y)>c' \\ c'(1-\epsilon_{1,k})p_0^*(y) &\textnormal{       for      } p_1^*(y)/p_0^*(y) \leq c',
\end{cases}
\end{dmath*} 
where $0\leq c'<c''\leq \infty$ are determined such that $q_{0,k}$ and $q_{1,k}$ are probability densities. Note that $c'=0$ if and only if $\epsilon_{1,k}=0$, and $c'' = \infty$ if and only if $\epsilon_{0,k}=0$. Let $b_k=(1-\epsilon_{1,k})/(1-\epsilon_{0,k})$. We then have
\begin{dmath}\label{eqn:LRLFD}
{\frac{q_{1,k}(y)}{q_{0,k}(y)} =
\begin{cases}
 b_kc' &\textnormal{for } p_1^*(y)/p_0^*(y) \leq c'\\
 b_k\cdot\frac{p_1^*(y)}{p_0^*(y)} &\textnormal{for } c'<\frac{p_1^*(y)}{p_0^*(y)}<c''\\
b_kc'' &\textnormal{for } p_1^*(y)/p_0^*(y) \geq c''.
\end{cases}}
\end{dmath} 
In \cite{Hub:65}, it was shown that the LFDs of a pair of uncertainty classes are the two distributions that give the largest error probability when using a likelihood-ratio test to tell them apart.

In the rest of this paper, for any random variable $Y$ with distributions drawn from a given pair of uncertainty classes, we let $l^*(Y)$ be the likelihood ratio $q_1(Y)/q_0(Y)$, where $q_0$ and $q_1$ are the respective densities of the LFDs of the given uncertainty classes. In addition, we use $l^*(Y=y)$ to denote the realization of $l^*(Y)$ when $Y=y$. A sequence $x_1, x_2, \ldots, x_n$ is denoted as $(x_i)_{i=1}^n$.

\section{Robust Learning in a Tandem Network}\label{section:robust}
When there is only a single agent, the minimax error $\inf_{\phi} P_1^{\text{SL}}(\phi)$ is achieved by setting $\phi$ to be a likelihood ratio test using the LFDs $(Q_{0,1},Q_{1,1})$ \cite{Hub:65}. A similar result is proven in \cite{VeeBasPoo:94} for a parallel network configuration. In this section, we show the same result for the tandem network. 
We do this by first showing that for any sequence of agent decision rules that consists of likelihood ratio tests between LFDs, the error probabilities \eqref{detection_prob} and \eqref{social_learning} are maximized when all the private agent observations are drawn from the corresponding LFDs. We then show that when agents' private observations are drawn from the respective LFDs, then likelihood ratio tests using LFDs minimize the error probabilities. We first state the following lemma proven in \cite{Hub:65}.

\begin{lemma}\label{lemma:Huber}
Suppose that the LFDs for $(\cP_0,\cP_1)$ are $(Q_0,Q_1)$. Then, for any random variable $Y$ with distributions 
$P_i\in \mathcal{P}_i$, where $i=0,1$, we have \begin{dmath*}P_0({l^*(Y)>t})\leq Q_0({l^*(Y)>t}) \leq Q_1({l^*(Y)>t}) \leq P_1({l^*(Y)>t}).
\end{dmath*}
\end{lemma}

We can now present our first result. For all $k \geq 1$, let $(Q_{0,k},Q_{1,k})$ be the LFDs for $(\cP_{0,k},\cP_{1.k})$, and $Q_i^{(N)}=Q_{i,1}\times Q_{i,2}\times ... \times Q_{i,N}$ for $i=0,1$.

\begin{theorem}\label{theorem:LFD_tandem}
Let $\phi^{(N)}$ be any sequence of monotone likelihood ratio tests based on the LFDs $(Q_0\tn{N},Q_1\tn{N})$ for a tandem network. Then for all $(P_0^{(N)},P_1^{(N)})\in  \mathcal{P}_0^{(N)}\times\mathcal{P}_1^{(N)}$, we have
\begin{dmath}\label{ineq:PF_Q}
P_{F,N}(\phi^{(N)},Q_0^{(N)})\geq P_{F,N}(\phi^{(N)},P_0^{(N)}),
\end{dmath}
and
\begin{dmath}\label{ineq:PM_Q}
P_{M,N}(\phi^{(N)},Q_1^{(N)})\geq P_{M,N}(\phi^{(N)},P_1^{(N)}).
\end{dmath}
\end{theorem}
\begin{IEEEproof}
We proceed by mathematical induction on $N$. From Lemma \ref{lemma:Huber}, the theorem holds for $N=1$. We now assume that it holds for $N<i$. We also make use of the following two lemmas, the first of which is proved in Appendix \ref{proof:lemma:Q1>Q0}, while the second is shown in \cite{VeeBasPoo:94}.
\begin{lemma}\label{lemma:Q1>Q0}
For any $N$, $Q_1\tn{N}(U_N=1)\geq Q_0\tn{N}(U_N=1)$ and $Q_1\tn{N}(U_N=0)\leq Q_0\tn{N}(U_N=0)$.
\end{lemma}


\begin{lemma}\label{lemma_larger}
Let $Z_1$ and $Z_2$ be non-negative, independent random variables. If for $k=1,2$, we have
\begin{dmath}\label{ineq:stochastically_larger}
{F(Z_k>t)\geq G(Z_k>t),\quad\forall\:t\geq 0,}
\end{dmath}
then
\begin{dmath*}
{F(Z_1Z_2>t)\geq G(Z_1Z_2>t),\quad\forall\:t\geq 0.}
\end{dmath*}
\end{lemma}  
If \eqref{ineq:stochastically_larger} holds, we say that $Z_k$ is stochastically larger under $F$ than $G$. Since the observation of agent $k$ is independent from the decision it receives, for $i=0,1$ we have 
\begin{dmath*}
Q_i\tn{k}(l^*(Y_k)> t)=Q_{i,k}(l^*{(Y_k)> t}).
\end{dmath*}
From Lemma \ref{lemma:Huber}, $l^*(Y_k)$ is stochastically larger under $Q_0\tn{k}$ than under any other distribution $P_0\tn{k}\in \cP_0^k$. To show the same for $l^*(U_{k-1})$, we obtain from Lemma \ref{lemma:Q1>Q0} that
\begin{dmath*}
l^*(U_{k-1}=1)=\frac{Q_1\tn{k-1}(U_{k-1}=1)}{Q_0\tn{k-1}(U_{k-1}=1)}
\geq 1
\geq \frac{1-Q_1\tn{k-1}(U_{k-1}=1)}{1-Q_0\tn{k-1}(U_{k-1}=1)}
\geq l^*({U_{k-1}=0}).
\end{dmath*}
For any $l^*(U_{k-1}=0)<t<l^*({U_{k-1}=1})$,
\begin{dmath*}
Q_0\tn{k}(l^*(U_{k-1})>t)=Q_0\tn{k}({U_{k-1}=1})
\geq P_0\tn{k}({U_{k-1}=1})
=P_0\tn{k}(l^*({U_{k-1})>t}),
\end{dmath*}
where the inequality follows from the induction hypothesis. Note that $l^*(U_{k-1})$ only takes the two values $l^*(U_{k-1}=0)$ and $l^*(U_{k-1}=1)$, so for any $t$ not in between these values, the equality $Q_0\tn{k}(l^*(U_{k-1})>t)=P_0\tn{k}(l^*(U_{k-1})>t)$ is trivially true. Therefore, $l^*(U_{k-1})$ is stochastically larger under $Q_0\tn{k}$.

From Lemma \ref{lemma_larger}, the product of $l^*(Y_k)$ and $l^*(U_{k-1})$ is stochastically larger under $Q_0\tn{k}$ than under any other distribution $P_0\tn{k}$ as well. Therefore, we have
\begin{dmath*}
Q_0\tn{k}(U_k=1) = Q_0\tn{k}({l^*(U_{k-1},Y_k)>t_k})
 \geq  P_0\tn{k} ({l^*(U_{k-1},Y_k)>t_k}) 
=P_0({U_k=1}).
\end{dmath*}
The proof for the missed detection probability inequality \eqref{ineq:PM_Q} is similar, and the induction is complete. The theorem is now proved.
\end{IEEEproof}

\begin{theorem}\label{theorem:minimax}
Let $\phi_*\tn{N}$ be an optimal sequence of decision rules such that
\begin{dmath*}
\phi_*\tn{N} = \arg\min_{\phi\tn{N}} P_{e,N}(\phi\tn{N}, Q_0\tn{N}, Q_1\tn{N}).
\end{dmath*}
Then, $\phi_*\tn{N}$ minimizes $P_N^{\text{DD}}(\cdot)$ in \eqref{detection_prob}. Similarly, for each $k \geq 1$, define $\psi_{*,k}$ recursively as
\begin{dmath*}
\psi_{*,k} =  \arg\min_{\psi_k} P_{e,k}((\psi_*\tn{k-1},\psi_k), Q_0\tn{k}, Q_1\tn{k}),
\end{dmath*}
where $\psi_{*,0}$ is ignored. Then, $\psi_{*,k}$ minimizes ${P_k^{\text{SL}}(\cdot \mid  \psi_*\tn{k-1})}$ in \eqref{social_learning} for all $k\geq 1$. 

\end{theorem}
\begin{IEEEproof}
In \cite{Tsi:93}, it was shown that $\phi_*\tn{N}$ is a sequence of likelihood ratio tests based on $(Q_0\tn{N},Q_1\tn{N})$. Hence, for any sequence of decision rules $\phi\tn{N}$, we have from Theorem \ref{theorem:LFD_tandem},
\begin{dgroup*}
\dm{\sup_{(P_0\tn{N},P_1\tn{N})\in\mathcal{P}_0\tn{N}\times\mathcal{P}_1\tn{N}}P_{e,N}(\phi_*\tn{N}, P_0\tn{N}, P_1\tn{N})
= P_{e,N}(\phi_*\tn{N}, Q_0\tn{N}, Q_1\tn{N})}
\dm[eq:minimax_opt]{\leq P_{e,N}(\phi\tn{N}, Q_0\tn{N}, Q_1\tn{N})}
\dm{\leq \sup_{(P_0\tn{N},P_1\tn{N})\in\mathcal{P}_0\tn{N}\times\mathcal{P}_1\tn{N}}P_{e,N}(\phi\tn{N}, P_0\tn{N}, P_1\tn{N})}
\end{dgroup*}
where \eqref{eq:minimax_opt} follows from the theorem assumption. Therefore, the minimax error in the decentralized detection problem is equal to the minimum error when all the distributions of the private observations are equal to the LFDs. A similar argument holds for the social learning problem in the second part of this theorem. The proof is now complete.
\end{IEEEproof}

We remark that Theorem \ref{theorem:minimax} can be extended to general tree topologies. This is because in such a topology, the decisions received by each agent are mutually independent. Hence,
\begin{dmath*}
l^*(Y_k,U_{k_1},\dots,U_{k_m})=l^*(Y_k)\displaystyle\prod_{j=1}^ml^*(U_{k_j}),
\end{dmath*}
where $k_1,\dots,k_m$ are the agents that agent $k$ is receiving decisions from. The rest of the proof then proceeds similarly as that of Theorems \ref{theorem:LFD_tandem} and \ref{theorem:minimax}.

\section{Asymptotic Detection and Social Learning}\label{section:asymptotic}

Theorem \ref{theorem:minimax} shows that there is no loss in optimality in both the decentralized detection and social learning problems if agents in a tandem network are restricted to monotone likelihood ratio tests based on the LFDs. It however does not tell us the minimum minimax error probability achievable. In this section, we study the minimax error probability in long tandems under various technical assumptions in order to simplify the problem. In particular, we investigate the conditions under which the minimax error probability converges to zero as the number of agents increases.\footnote{In the case where the agents' contamination values for their uncertainty classes are zero, \cite{AceDahLob:08} calls this \emph{asymptotic learning}.} We first consider the case where every agent has identical uncertainty classes, and provide necessary and sufficient conditions for the minimax error probability to approach zero under both decentralized detection and social learning. We will then proceed to analyze social learning in long tandems where the contamination of the uncertainty class can differ. Finally, we study the achievable asymptotic minimax error probability when agents do not know their own positions in the tandem.

\subsection{Identical Uncertainty Classes}
In this subsection, we make the following assumption that the uncertainty classes of every agent are identical. 
\begin{assumption}\label{assumpt:identical}
For all $k\geq 1$, we have $\epsilon_{0,k}=\epsilon_0$, $\epsilon_{1,k}=\epsilon_1$, and the LFDs of each agent's uncertainty classes are $(Q_0, Q_1)$.
\end{assumption}
The following two results give necessary and sufficient conditions for the minimax error probability to approach zero under the decentralized detection and social learning frameworks, respectively.

\begin{theorem}[Decentralized detection]\label{theorem_asymptotic}
Suppose that Assumption \ref{assumpt:identical} holds, and that the decision rules for every agent are chosen so as to minimize $P_{N}^{\text{DD}}(\phi^{(N)})$ in \eqref{detection_prob}. Then $P_{N}^{\text{DD}}(\phi^{(N)}) \to 0$ as $N\to\infty$ if and only if at least one of the following is true:
\begin{enumerate}[1)]
\item $\epsilon_{0}=0$ and the log-likelihood ratio of $P_{1}^*$ versus $P_{0}^*$ is unbounded from above,
\item $\epsilon_{1}=0$ and the log-likelihood ratio of $P_{1}^*$ versus $P_{0}^*$ is unbounded from below.
\end{enumerate}
\end{theorem} 
\begin{IEEEproof}
See Appendix \ref{proof:theorem_asymptotic}.
\end{IEEEproof}

\begin{theorem}[Social learning]\label{theorem:asymptotic_social}
Suppose that Assumption \ref{assumpt:identical} holds, and that the decision rule for each agent is chosen sequentially so as to minimize $P_k^{\text{SL}}(\phi_k \mid  \phi\tn{k-1})$ in \eqref{social_learning}. Then $P_k^{\text{SL}}(\phi_k \mid  \phi\tn{k-1}) \to 0$ as $k\to\infty$ if and only if $\epsilon_0=\epsilon_1=0$ and the log-likelihood ratio of $P_1^*$ versus $P_0^*$ is unbounded.
\end{theorem}
\begin{IEEEproof}
It was demonstrated in Proposition 3 of \cite{PapAth:92} that using social learning decision rules, when $\epsilon_0=\epsilon_1=0$, the error probability in a tandem network where all agents have the same observation distributions will converge to zero if and only if the log-likelihood ratios between the two probability distributions is unbounded from both above and below. The sufficiency of the given condition then follows immediately. To show that it is also necessary, observe that if $\epsilon_0>0$ or $\epsilon_1>0$, then the log-likelihood ratio of $Q_1$ versus $Q_0$ is bounded from either above or below respectively, and Theorem \ref{theorem:minimax} and Proposition 3 of \cite{PapAth:92} implies that $P_N^{\text{DD}}(\phi^{(N)})$ is bounded away from zero. The proof is now complete.
\end{IEEEproof}
From the theorems above, it can be seen that for asymptotic learning to occur in the social learning case, it is necessary that both the distributions be uncontaminated. In the decentralized detection case, it is only necessary that one of the distributions be uncontaminated.

\subsection{Varying uncertainty classes for social learning}
In this subsection, we relax the assumption of identical uncertainty classes for all agents in the previous subsection, and study the effect of varying contamination values on the asymptotic error probability in the social learning framework. We make the following assumptions.
\begin{assumption}\label{assumpt:varying}
We have
\begin{enumerate}[(i)]
	\item\label{it:LLR} the log-likelihood ratio of $P_1^*$ versus $P_0^*$ is unbounded; and
	\item\label{it:epsilon} each agent $k\geq 1$ knows its own contamination values $\epsilon_{i,k}$, for $i=0,1$, as well as those of its predecessors, and its position in the tandem network.
\end{enumerate}
\end{assumption}
Assumption \ref{assumpt:varying}\eqref{it:LLR} is necessary because otherwise learning the true hypothesis is not possible under a social learning framework even if all the contamination values are zero, as shown in Theorem \ref{theorem:asymptotic_social}. We will show that under these assumptions, learning the true hypothesis happens if there exist infinite subsequences $(\epsilon_{0,k_n})_{n\geq 1}$ and $(\epsilon_{1,j_n})_{n\geq 1}$ (which may potentially be distinct) that converge to zero as $n$ increases.

We first observe that under the social learning framework, agents minimize their local maximum error probability. This implies that regardless of the values of $\epsilon_{0,k}$ and $\epsilon_{1,k}$, the minimax error probability of each agent $k$ is non-increasing in $k$. This is because any agent can simply pass on the decision of the previous agent if no other decision rule leads to a decrease in minimax error probability.

For ease of notation, we let $Q_{F,k}=P_{F,k}(\phi^{(k)},Q_0^{(k)})$ and $Q_{M,k}=P_{M,k}(\phi^{(k)},Q_0^{(k)})$ be the LFD false alarm and miss detection probabilities of agent $k$ respectively, where $\phi^{(k)}$ is the sequence of optimal social learning decision rules that minimizes $P_{k}^{\text{SL}}(\phi_k \mid  \phi\tn{k-1})$ in \eqref{social_learning}. From \cite{PapAth:92}, it can be shown that the decision rule used by agent $k$ is of the form
\begin{dmath}\label{varying_rule}
u_k=\begin{cases}
0\:\:\:\:\:\:\:\textnormal{ if } l^*(Y_{k})<\frac{\pi_0Q_{F,{k-1}}}{\pi_1(1-Q_{M,{k-1}})}
\\1\:\:\:\:\:\:\:\textnormal{ if }l^*(Y_k)\geq \frac{\pi_0(1-Q_{F,{k-1}})}{\pi_1Q_{M,{k-1}}}
\\u_{k-1}\textnormal{ if }\frac{\pi_0 Q_{F,{k-1}}}{\pi_1(1-Q_{M,{k-1}})}\leq l^*(Y_k)<\frac{\pi_0(1-Q_{F,{k-1}})}{\pi_1Q_{M,{k-1}}}.
\end{cases}
\end{dmath}

\begin{theorem}\label{theorem:varying}
Suppose that Assumption \ref{assumpt:varying} holds, and that each agent $k$ in a tandem network chooses decision rule $\phi_k$ to minimize $P_k^{\text{SL}}(\phi_k \mid  \phi\tn{k-1})$ in \eqref{social_learning}. Then $ {P_k^{\text{SL}}(\phi_k \mid  \phi\tn{k-1})} \rightarrow 0$ as $k\rightarrow\infty$ if and only if there exist infinite subsequences $\epsilon_{0,k_n} \to 0$ and $\epsilon_{1,j_n} \to 0$ as $n\to\infty$.
\end{theorem}
\begin{IEEEproof}
See Appendix \ref{proof:theorem:varying}.
\end{IEEEproof}
We observe that agents in a tandem network in the social learning framework have total error probabilities at least that of agents adopting decentralized detection rules. Hence, Theorem \ref{theorem:varying} also provides a sufficient condition for the minimax error probability \eqref{detection_prob} under a decentralized detection framework to converge to zero.
%

\subsection{Unknown agent positions}\label{section:unknown}

In a social network, users have to make their decisions not knowing how many hops information has been propagated from a source node. We model this in a tandem network by assuming that each agent has no knowledge of its position in the network. We make the following assumption, in addition to Assumption \ref{assumpt:identical}, in this subsection.
\begin{assumption}\label{assumpt:unknown}
Every agent $k > 1$ uses the same decision rule.
\end{assumption}
Except for agent 1 (which knows its position in the tandem because it does not receive any preceding messages), every other agent in the tandem does not know its own position, and has access to exactly the same information when it comes to choosing a decision rule, which is based solely on the nominal distributions, $P_0^*$ and $P_1^*$, as well as the contamination values $\epsilon_0$ and $\epsilon_1$. Therefore, it is natural to make Assumption \ref{assumpt:unknown}. 

Because of Assumption \ref{assumpt:unknown}, any sequence of $k$ agent decision rules has the form $\phi\tn{k} = (\phi_1, \phi^{k-1})$, where $\phi_1$ is the decision rule used by the first agent, and $\phi$ is the decision rule used by every other agent with $\phi^{k-1}= (\phi, \ldots, \phi)$ consisting of $k-1$ copies of $\phi$. For simplicity, and by abusing notation, we replace $\phi\tn{k}$ in our notations by $(\phi_1,\phi)$ in the sequel.

In the following, we consider two different scenarios.  

\subsubsection{Minimizing asymptotic error}
We consider the case where agents are collaborating to minimize the asymptotic error. This might occur when there is a chain of agents trying to relay some information to a fusion center, but each agent is unsure of how many other agents there are between itself and the fusion center.
For a given decision rule $\phi$, the asymptotic maximum error probability is given by
\begin{dmath}
P_\infty^{\text{DD}}(\phi)=\lim_{k\rightarrow\infty}P_k^{\text{DD}}(\phi_1,\phi), \label{eq:asymptotic_error}
\end{dmath}
where we have implicitly assumed that $P_\infty^{\text{DD}}$ does not depend on $\phi_1$. This assumption is valid, as shown in the next theorem, which also provides a characterization for the optimal $\phi$ which obtains the asymptotic minimax error probability, defined as
\begin{dmath*}
\inf_\phi P_\infty^{\text{DD}}(\phi).
\end{dmath*} 
In the previous sections, we had no need to consider randomized decision rules. This is because under the decentralized detection framework, the final error probability under a randomized sequence of decision rules is no less than the minimum final error probability of each of the respective deterministic sequences of decision rules. Similarly, under the social learning framework, the error probability of an agent using a randomized decision rule is no less than the minimum error probability under each of the deterministic rules. However, this property does not hold for the asymptotic maximum error probability. A randomized version of two deterministic decision rules may yield a lower asymptotic maximum error probability than either of the two deterministic rules. Thus, we introduce the randomized likelihood ratio test:
\begin{dmath}\label{ex_LRT}
u_k=\begin{cases}
0\textnormal{ if }l^*(Y_k)<t_1
\\A\textnormal{ if }l^*(Y_k)=t_1
\\u_{k-1}\textnormal{ if }t_1<l^*(Y_k)<t_0
\\B\textnormal{ if }l^*(Y_k)=t_0
\\1\textnormal{ if }l^*(Y_k)> t_0,
\end{cases}
\end{dmath}
where $A$ is equal to 0 with probability $p$ and equal to $u_{k-1}$ with probability $1-p$, $B$ is equal to $u_{k-1}$ with probability $q$ and equal to $1$ with probability $q$, and $t_1$, $t_0$, $p$ and $q$ are constants to be determined.

We will now show how to obtain the minimax asymptotic error probability. To do so, we start with the following two lemmas.
\begin{lemma}\label{lemma:recurrence}
Suppose that Assumptions \ref{assumpt:identical} and \ref{assumpt:unknown} hold. Then,
\begin{dmath*}
\lim_{k\to\infty} P_{e,k}((\phi_1,\phi), Q_0^k, Q_1^k)
=\frac{\pi_0Q_0(\phi(Y_1,0)=1)}{Q_0(\phi(Y_1,0)=1)+Q_0(\phi(Y_1,1)=0)}+\frac{\pi_1Q_1(\phi(Y_1,1)=0)}{Q_1(\phi(Y_1,1)=0)+Q_1(\phi(Y_1,0)=1)}.
\end{dmath*}
\end{lemma}
\begin{IEEEproof}
We have the two following recurrence relations:
\begin{dgroup*}
\dm{P_{F,k}((\phi_1,\phi), Q_0^k)
=Q_0^k({U_k=1})
={Q_0^{k-1}({U_{k-1}=0})\cdot Q_0({{U_k=1}\mid {U_{k-1}=0}})}+Q_0^{k-1}({U_{k-1}=1})\cdot Q_0({U_k=1\mid U_{k-1}=0})
=(1-Q_0^{k-1}({U_{k-1}=1}))\cdot Q_0({\phi(Y_k,0)=1})+Q_0^{k-1}({U_{k-1}=1})\cdot Q_0({\phi(Y_k,1)=1})}
\dm[eq:recurrence]{=P_{F,k-1}((\phi_1,\phi), Q_0^{k-1})\cdot [Q_0({\phi(Y_1,1)=1})-Q_0({\phi(Y_1,0)=1})]\\+Q_0({\phi(Y_1,0)=1}),}
\end{dgroup*}
and
\begin{dgroup*}
\dm{P_{M,k}((\phi_1,\phi), Q_0^k)
=Q_1^k({U_k=0})
={Q_1^{k-1}({U_{k-1}=0})\cdot Q_1({{U_k=0}\mid {U_{k-1}=0}})}+Q_1^{k-1}({U_{k-1}=1})\cdot Q_1({U_k=0\mid U_{k-1}=1})
=Q_1^{k-1}({U_{k-1}=0})\cdot Q_1({\phi(Y_k,0)=0})+(1-Q_1^{k-1}({U_{k-1}=0}))\cdot Q_0({\phi(Y_k,1)=0})}
\dm[eq:recurrence2]{=P_{M,k-1}((\phi_1,\phi), Q_1^{k-1})\cdot [Q_1({\phi(Y_1,1)=1})-Q_1({\phi(Y_1,0)=1})]\\+Q_1({\phi(Y_1,0)=0}),}
\end{dgroup*}
The first recurrence relation converges linearly to\begin{dmath}\frac{Q_0(\phi(Y_1,0)=1)}{Q_0(\phi(Y_1,0)=1)+Q_0(\phi(Y_1,1)=0)},
\end{dmath}
and the second recurrence relation converges linearly to
\begin{dmath}\frac{\pi_1Q_1(\phi(Y_1,1)=0)}{Q_1(\phi(Y_1,1)=0)+Q_1(\phi(Y_1,0)=1)}.
\end{dmath}
Hence, the proof is complete.
\end{IEEEproof}

Next, we will show that if the observation distributions are all drawn from the LFDs, then the decision rule that minimizes that asymptotic error probability is a randomized likelihood ratio test between $Q_0$ and $Q_1$.

\begin{lemma}\label{lemma:8}
Suppose that Assumptions \ref{assumpt:identical} and \ref{assumpt:unknown} hold. For any $\phi_1$, let 
\begin{dmath*}\phi_*=\arg\min_\phi \lim_{k\to\infty}P_{e,k}((\phi_1,\phi), Q_0^k, Q_1^k).
\end{dmath*}
Then there is no loss in optimality if $\phi_*$ is restricted to be a randomized likelihood ratio test between $Q_0$ and $Q_1$.
\end{lemma}
\begin{IEEEproof}
Assume that $\phi_*$ is not a randomized likelihood ratio test between $Q_0$ and $Q_1$. For any agent $k\geq 2$, consider a randomized likelihood ratio test $\phi'$ in the form presented in \eqref{ex_LRT}, where $t_0,t_1,p,q$ are chosen such that
\begin{dmath*}
Q_0({\phi'(Y_k,0)=1})=Q_0({l^*(Y_k)\geq t_0})=Q_0({\phi_*(Y_k,0)=1})
\end{dmath*}
\begin{dmath*}
Q_1({\phi'(Y_k,1)=0})=Q_1({l^*(Y_k)< t_1})=Q_1({\phi_*(Y_k,1)=0}).
\end{dmath*}
From the Neyman-Pearson lemma, we then have
\begin{dmath*}
Q_1({\phi'(Y_k,0)=0})=Q_1({l^*(Y_k)< t_0})\leq Q_1({\phi_*(Y_k,0)=0})
\end{dmath*}
\begin{dmath*}
Q_0({\phi'(Y_k,1)=1})=Q_0({l^*(Y_k)< t_1})\leq Q_0({\phi_*(Y_k,1)=1}).
\end{dmath*}
Hence, from \eqref{eq:recurrence} and \eqref{eq:recurrence2}, it is clear that for any $\phi_1$ and any $k\geq 1$, we have 
\begin{dmath*}
P_{F,k}((\phi_1,\phi_*), Q_0^k, Q_1^k)\geq P_{F,k}((\phi_1,\phi'), Q_0^k, Q_1^k),
\end{dmath*}
and
\begin{dmath*}
P_{M,k}((\phi_1,\phi_*), Q_0^k, Q_1^k)\geq P_{M,k}((\phi_1,\phi'), Q_0^k, Q_1^k),
\end{dmath*}
and hence
\begin{dmath*}
P_{e,k}((\phi_1,\phi_*), Q_0^k, Q_1^k)\geq P_{e,k}((\phi_1,\phi'), Q_0^k, Q_1^k).
\end{dmath*}
By choosing suitable values of $t_0,t_1,p,q$, we can set $Q_0({\phi(Y_k,0)=1})$ and $Q_0({\phi(Y_k,1)=1})$ to any value between 0 and 1. Similarly, we can also set $Q_1({\phi(Y_k,1)=0})$ and $Q_1({\phi(Y_k,0)=0})$ to any value between 0 and 1. Hence, 
\begin{dmath*}
\min_\phi \lim_{k\to\infty}P_{e,k}((\phi_1,\phi), Q_0^k, Q_1^k)
\end{dmath*}
is attainable, and the decision rule used to attain it can be assumed to be in the form of a randomized likelihood ratio test.
\end{IEEEproof}
The proof of Lemma~\ref{lemma:8} shows that the asymptotic minimax error probability is attainable even when the likelihood ratio of $Y_k$ is not continuous. To avoid cumbersome notation, for the rest of the paper, we will assume that the likelihood ratio of $Y_k$ is continuous. It is easy to extend our results if this is not the case.

We can now prove the following theorem, which provides an expression for the minimax asymptotic error.

\begin{theorem}\label{theorem:unknown_decentralized}
Suppose that Assumptions \ref{assumpt:identical} and \ref{assumpt:unknown} hold, and let $\phi_*$ be the randomized likelihood ratio test such that
\begin{dmath*}
\phi_*=\arg\min_\phi \lim_{k\rightarrow\infty}P_{e,k}((\phi_1,\phi), Q_0^k, Q_1^k).
\end{dmath*}
Then,
\begin{dmath*}
\inf_\phi P_\infty^{\text{DD}}(\phi)=\lim_{k\rightarrow\infty}P_{e,k}((\phi_1,\phi_*), Q_0^k, Q_1^k).
\end{dmath*}
\end{theorem}
\begin{IEEEproof}
From Theorem \ref{theorem:LFD_tandem}, we have
\begin{dmath*}
\lim_{k\rightarrow\infty}P_{e,k}((\phi_1,\phi_*), Q_0^k, Q_1^k)=\lim_{k\rightarrow\infty}\sup_{(P_0\tn{k},P_1\tn{k})\in\mathcal{P}_0^{k}\times\mathcal{P}_1^{k}}P_{e,k}((\phi_1,\phi_*), P_0\tn{k}, P_1\tn{k})
=P_\infty^{\text{DD}}(\phi_*).
\end{dmath*}
By the theorem assumption, for any decision rule $\phi$ we have
\begin{dmath*}
\lim_{k\rightarrow\infty}P_{e,k}((\phi_1,\phi_*), Q_0^k, Q_1^k)\leq\lim_{k\rightarrow\infty}P_{e,k}((\phi_1,\phi), Q_0^k, Q_1^k).
\end{dmath*}
Hence, for any decision rule $\phi$, we have
\begin{dmath*}
P_\infty^{\text{DD}}(\phi)=\lim_{k\rightarrow\infty}\sup_{P_0^{(k)}\in\mathcal{P}_0^k,P_1^{(k)}\in\mathcal{P}_1^k}P_{e,k}((\phi_1,\phi), P_0\tn{k}, P_1\tn{k})
\geq \lim_{k\rightarrow\infty}P_{e,k}((\phi_1,\phi), Q_0^k, Q_1^k)
\geq \lim_{k\rightarrow\infty}P_{e,k}((\phi_1,\phi_*), Q_0^k, Q_1^k)
=P_\infty^{\text{DD}}(\phi_*).
\end{dmath*}
The proof is now complete.
\end{IEEEproof}
Theorem \ref{theorem:unknown_decentralized} states that each agent should find the decision rule to optimize the asymptotic minimax error as if its observations were distributed according the the LFDs of the uncertainty class. This is consistent with our results when agents do know their positions (Theorem \ref{theorem:LFD_tandem}). However, the exact threshold values for $\phi_*$ are difficult to compute in general, but can be found using numerical methods. Together with Lemma \ref{lemma:recurrence}, since the uncertainty classes $\mathcal{P}_{0,k}$ and $\mathcal{P}_{1,k}$ are disjoint for all $k$, Theorem \ref{theorem:unknown_decentralized} shows that asymptotically learning the true hypothesis is impossible when agents do not know their own positions and also provides an expression for the asymptotic minimax error.

\subsubsection{Minimizing error of current agent}
We now assume that every agent is acting to minimize its local minimax error probability, and that each agent past the first does not know which position it is in. This is true in general for most social networks, where it is difficult to find the root of any information spread. Hence, users typically would not know how many hops information has been propagated from its source.

We find the decision rule to minimize the maximum error probability by allowing each agent to consider the maximum error for every possible position it might be in, then finding the decision rule that minimizes this value. Like in the previous subsection, we also assume that Assumptions \ref{assumpt:identical} and \ref{assumpt:unknown} are in effect. More specifically, for each agent $k\geq 2$, we wish to find $\phi$ that minimizes
\begin{dmath}
P_{\max}^{\text{SL}}(\phi) =\sup_{k\geq 2} P_k^{\text{SL}}({\phi \mid(\phi_1,\phi)}),\label{eq:phi}
\end{dmath}
where $\phi_1$ is the optimal decision rule that minimizes $P_1^{\text{SL}}(\cdot)$. We will show that for the optimal decision rule $\phi$, the maximum error probability occurs either in the second position or at the asymptotic limit, as defined in \eqref{eq:asymptotic_error}. 

\begin{theorem}\label{theorem:unknown_social}
Suppose that Assumptions \ref{assumpt:identical} and \ref{assumpt:unknown} hold. Then $\phi_* = \arg\min_{\phi} P_{\max}^{\textrm{SL}}(\phi)$ is a randomized likelihood ratio test between the LFDs $Q_0$ and $Q_1$, and 
\begin{dmath*}
P_{\max}^{\textrm{SL}}(\phi_*) = \max\left\{P_2^{\text{SL}}({\phi_* \mid \phi_1)}, P_\infty^{\text{DD}}(\phi_*)\right\},
\end{dmath*} 
where $P_\infty^{\text{DD}}(\phi_*)$ is as defined in \eqref{eq:asymptotic_error}.
\end{theorem}
\begin{IEEEproof}
See Appendix \ref{proof:unknown_social}.
\end{IEEEproof}

Finding an analytical form for $\phi_*$ in Theorem \ref{theorem:unknown_social} is difficult. However, as $\phi_*$ is known to be a randomized likelihood ratio test of $Q_0$ and $Q_1$, this can be done numerically by minimizing $\max\{P_2^{\text{SL}}(\phi_* \mid \phi_1), P_\infty^{\text{DD}}(\phi_*)\}$ with respect to the thresholds for $\phi_*$.

\section{Numerical Results}\label{section:numeric}
In this section, we provide numerical results to illustrate part of our theoretical contributions in Section \ref{section:unknown}, which shows that even if an agent $k$'s position is unknown, where $k\geq 2$, its optimal decision rule $\phi_*$ can be chosen to be a randomized likelihood ratio test between the LFDs $Q_0$ and $Q_1$. This is true whether the agent is collaborating with others to minimize the asymptotic error (decentralized detection) or is trying to minimize its own error probability (social learning). We have shown this in Theorems \ref{theorem:unknown_decentralized} and \ref{theorem:unknown_social} respectively. The form of this randomized likelihood ratio test $\phi_*$ is given in \eqref{ex_LRT}. Now, we can rewrite this in terms of an optimzation problem.

First, given the nominal distributions $P_0^*$ and $P_1^*$, as well as the contamination values $\epsilon_0$ and $\epsilon_1$, we can use a binary search to compute $c'$ and $c''$. From the definition of the LFDs, we know that the range of thresholds we have to optimize over is bounded between $bc'$ and $bc''$, where $b=\displaystyle\frac{1-\epsilon_1}{1-\epsilon_0}$. Then, we can derive the LFDs $Q_0$ and $Q_1$ and obtain
\begin{dmath*}
{Q_0({\phi_*(Y,1)=1})}={Q_0({l^*(Y)>t_1})}+{(1-p)Q_0({l^*(Y)=t_1})},
\end{dmath*}
and
\begin{dmath*}	
{Q_0({\phi_*(Y,0)=1})}={Q_0({l^*(Y)>t_0})}+{(1-q)Q_0({l^*(Y)=t_0})}.
\end{dmath*}
Similarly, we have
\begin{dmath*}
Q_1({\phi_*(Y,1)=0})=Q_1({l^*(Y)<t_1})+pQ_1({l^*(Y)=t_1})
\end{dmath*}
and
\begin{dmath*}
Q_1({\phi_*(Y,0)=0})=Q_1({l^*(Y)<t_0})+qQ_1({l^*(Y)=t_0}).
\end{dmath*}

In the case where agents collaborate to minimize the asymptotic maximum error $P_\infty^{\text{DD}}$, we minimize the expression in Lemma \ref{lemma:recurrence} to obtain the optimal decision rule $\phi_*$. From Theorem \ref{theorem:unknown_decentralized}, since $\phi_*$ is a randomized likelihood ratio test, we can perform the optimization over $t_1$, $t_0$, $p$ and $q$ in \eqref{ex_LRT}.

Similarly, in the case where agents minimize their local maximum error probability $P_{\max}^{\text{SL}}$, we minimize the expression in Theorem \ref{theorem:unknown_social} over $t_1$, $t_0$, $p$ and $q$.

We now present a numerical example using the exponential distribution. First, we fix $P_0^*$ as an exponential distribution with mean 1 and let $P_1^*$ be an exponential distribution with mean 2. We set $\epsilon_0=\epsilon_1=0.01$, and $\pi_0=\pi_1=0.5$. Then, the optimal decision rule $\phi_1$ for agent 1 has the form:
\begin{dmath*}
u_1=\begin{cases}
0\textnormal{ if }l^*(Y_1)<1
\\1\textnormal{ if }l^*(Y_1)\geq 1.
\end{cases}
\end{dmath*} Note that for this case, if $t_1\neq c'$ and $t_0\neq c''$, then it does not matter what $p$ and $q$ are. This is because $Q_0(l^*(Y_k)=x)$ and $Q_1(l^*(Y_k)=x)$ are both zero unless $x=c'$ or $x=c''$ when the nominal distributions are both continuous. We plot $P_k^{\text{SL}}(\phi\mid (\phi_1,\phi))$ against $k$ when $\phi = \phi_A$ and $\phi_B$, which are randomized likelihood ratio tests of the form \eqref{ex_LRT}. Both rules have $t_1=c'$ and $p=1$, but $\phi_A$ has $t_0=5$ and $\phi_B$ has $t_0=1.1$.

\begin{figure}[ht]
  \centering
			\psfrag{P1}[][]{{$\phi_A$}}
		\psfrag{P2}[][]{{$\phi_B$}}
		\psfrag{j}[][]{Position $k$}
		\psfrag{e}[][]{$P_k^{\text{SL}}$}
	\includegraphics[width=0.49\textwidth]{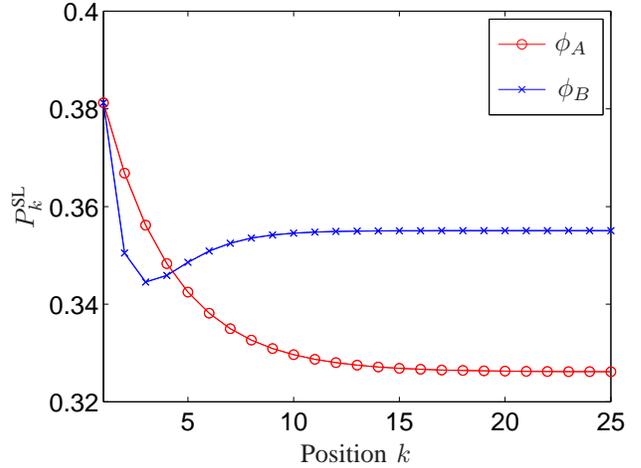}

	\caption{Comparison of decision rules.}
\label{figure:rules}
\end{figure}

Figure \ref{figure:rules} shows that the total error probability is decreasing over $k$ for $\phi_A$. Hence, the maximum error probability occurs when $k=2$. For $\phi_B$, the maximum occurs at the asymptotic limit $k\to\infty$. This is in line with our conclusion in Theorem \ref{theorem:unknown_social}.

Next, we let $P_0^*$ be an exponential distribution with mean 1 and $P_1^*$ be an exponential distribution with variable mean. We let $\epsilon_0=\epsilon_1=0.01$. We denote
\begin{dmath*}
\phi_*^{\text{SL}}=\arg\min_\phi P_{\max}^{\text{SL}}(\phi),
\end{dmath*}
and
\begin{dmath*}
\phi_*^{\text{DD}}=\arg\min_\phi P_{\infty}^{\text{DD}}(\phi),
\end{dmath*}
where $P_{\max}^{\text{SL}}(\phi)$ and $P_{\infty}^{\text{DD}}(\phi)$ are defined as in \eqref{eq:asymptotic_error} and \eqref{eq:phi} respectively.

\begin{figure}[ht]
  \centering
		\psfrag{DDDDD1}[cl][cl]{$\phi^{\text{SL}}_{\max} (\phi_*^{\text{SL}})$}
		\psfrag{DDDDD2}[cl][cl]{$\phi^{\text{DD}}_{\infty} (\phi_*^{\text{DD}})$}
		\psfrag{p1}[][]{Mean of $P_1^*$}
	\includegraphics[width=0.49\textwidth]{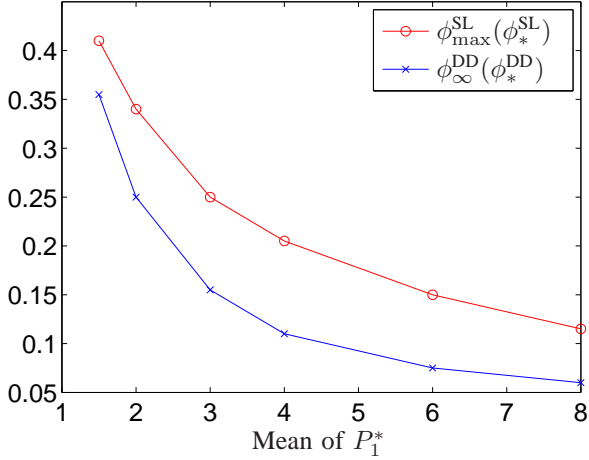}
	\caption{Error probability versus mean of $P_1^*$.}
\label{figure:mean}
\end{figure}

Figure \ref{figure:mean} shows that as the mean of $P_1^*$ increases, both $P_{\max}^{\text{SL}}(\phi_*^{\text{SL}})$ and $P_{\infty}^{\text{DD}}(\phi_*^{\text{DD}})$ decrease. This is intuitive as the Kullback-Leibler divergence of $P_1^*$ from $P_0^*$ increases as the mean of $P_1^*$ increases. As the nominal distributions become easier to differentiate, the asymptotic error probability decreases.

\begin{figure}[ht]
  \centering
		\psfrag{PSLPS}[cl][cl]{$\phi^{\text{SL}}_{\max} (\phi_*^{\text{SL}})$}
		\psfrag{PDDPD}[cl][cl]{$\phi^{\text{DD}}_{\infty} (\phi_*^{\text{DD}})$}
		\psfrag{C}[][]{Contamination ($\epsilon$)}
	\includegraphics[width=0.49\textwidth]{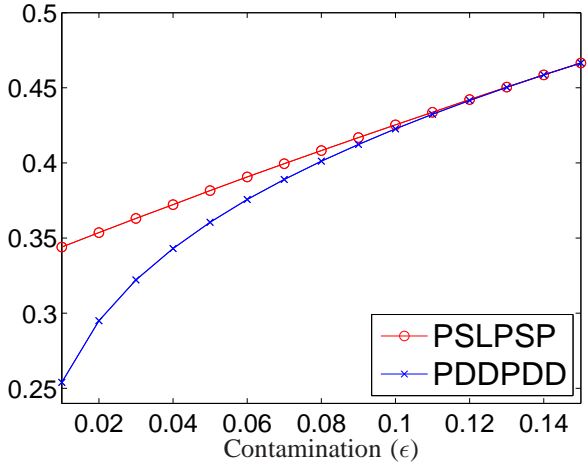}
	\caption{Error probability versus contamination value.}
\label{figure:epsilon}
\end{figure}

Figure \ref{figure:epsilon} shows that both $P_{\max}^{\text{SL}}(\phi_*^{\text{SL}})$ and $P_{\infty}^{\text{DD}}(\phi_*^{\text{DD}})$ increases as $\epsilon_0=\epsilon_1$ increases. For this graph, $P_0^*$ and $P_1^*$ are kept constant as exponential distributions with means 1 and 2 respectively. As expected, as uncertainty increases, so does the asymptotic error. When the uncertainty is large enough, both $P_{\max}^{\text{SL}}(\phi_*^{\text{SL}})$ and $P_{\infty}^{\text{DD}}(\phi_*^{\text{DD}})$ converge towards 0.5. Furthermore, as uncertainty increases, the gap between $P_{\max}^{\text{SL}}(\phi_*^{\text{SL}})$ and $P_{\infty}^{\text{DD}}(\phi_*^{\text{DD}})$ decreases. In a network with a lot of uncertainty, there is not much incentive in trying to get agents to collaborate, as agents selfishly trying to minimize their own error probability leads to similar error performance.

\section{Conclusion}\label{section:conclusion}
We have shown that in a tandem network where agents' observation distributions are not known exactly, and belong to uncertainty classes, the minimax error probability is obtained by assuming that each observation is distributed according to the LFDs of the uncertainty classes. In the case where agents know their positions in the tandem network, asymptotically learning the true hypothesis is in general impossible when the uncertainty classes have sizes bounded away from one, even when the log likelihood ratio of the nominal distributions of the uncertainty classes is unbounded. To achieve asymptotic learning of the true hypothesis in social learning, we require the additional condition that the uncertainty classes' contamination values decay over the tandem network. In the case where agents do not know their positions in the tandem, asymptotic learning of the true hypothesis becomes impossible even if contamination values are zero. We characterized the minimax error performance in this case, which provided a way to determine the optimal decision rules for the agents.

In this work, we have restricted our attention to the tandem network. It would be of interest to extend some of our results to tree networks, and even to loopy general graphs. Another future research direction would be to consider the robust detection problem with more than two hypotheses. A possible approach to this problem could be to focus on the LFDs on each possible pair of uncertainty class. Lastly, for the problem where each agent does not know its position in the network, we could instead consider each agent having partial knowledge of his position in the network, and find conditions under which learning the true hypothesis asymptotically is possible. 

\appendices
\section{Proofs of Main Results}\label{appendix}
\subsection{Proof of Lemma \ref{lemma:Q1>Q0}}\label{proof:lemma:Q1>Q0}
We will prove this lemma using mathematical induction on $N$. The likelihood ratio test for agent $k\geq 1$ is of the form 
\begin{dmath*}
 U_i=\begin{cases}
1 &\text{if } l^*(Y_k,U_{k-1})>t_k,\\
0 &\text{otherwise},
\end{cases}
\end{dmath*}
where $t_k$ is some chosen threshold. From Lemma \ref{lemma:Huber}, we have
\begin{dmath*}
Q_1\tn{1}(U_{1}=1)= Q_1\tn{1}({l^*(Y_1)\geq t_1})
\geq Q_0\tn{1}({l^*(Y_1)\geq t_1})
=Q_0\tn{1}({U_{1}=1}),
\end{dmath*}
so that the lemma holds for $N=1$.

Assume that the lemma is true for $N=k-1$. Each agent's observation is independent of the previous agent's decision. Hence, we have
\begin{dmath*}
l^*(Y_k,U_{k-1})=l^*(Y_k)l^*(U_{k-1}),
\end{dmath*}
and the likelihood ratio test can be rewritten in the form
\begin{dmath*}
 U_k=\begin{cases}
1 &\text{if } l^*(Y_k)\geq t_k^0,\\
U_{k-1} &\text{if } t_k^1\leq l^*(Y_k)<t_k^0\\
0 &\text{if } t_k^1> l^*(Y_k),
\end{cases}
\end{dmath*}
where $t_k^i=t_k/l^*(U_{k-1}=i)$. We obtain
\begin{dmath*}
Q_1\tn{k}({U_k=1})
=Q_{1,k}({U_k=1 \mid  U_{k-1}=1})Q_1\tn{k-1}({U_{k-1}=1})+Q_{1,k}({U_k=1\mid  U_{k-1}=0})Q_1\tn{k-1}({U_{k-1}=0})
=Q_{1,k}({l^*(Y_k)\geq t_k^1})Q_1\tn{k-1}({U_{k-1}=1})+Q_{1,k}({l^*(Y_k)\geq t_k^0})(1-Q_1\tn{k-1}({U_{k-1}=1})).
\end{dmath*}
From the induction hypothesis, we then have
\begin{dmath*}
Q_{1,k}({l^*(Y_k)\geq t_k^1})Q_1\tn{k-1}({U_{k-1}=1})+Q_{1,k}({l^*(Y_k)\geq t_k^0})(1-Q_1\tn{k-1}({U_{k-1}=1}))
\geq Q_{1,k}({l^*(Y_k)\geq t_k^1})Q_0\tn{k-1}({U_{k-1}=1})+Q_{1,k}({l^*(Y_k)\geq t_k^0})(1-Q_0\tn{k-1}({U_{k-1}=1}))
\geq Q_{0,k}({l^*(Y_k)\geq t_k^1})Q_0\tn{k-1}({U_{k-1}=1})+Q_{0,k}({l^*(Y_k)\geq t_k^0})(1-Q_0\tn{k-1}({U_{k-1}=1}))
=Q_0\tn{k}({U_k=1}),
\end{dmath*}
where the second inequality follows from Lemma \ref{lemma:Huber}. The proof of the lemma is now complete.

\subsection{Proof of Theorem \ref{theorem_asymptotic}}\label{proof:theorem_asymptotic}
We consider three separate cases, depending on whether $\epsilon_0$ or $\epsilon_1$ is nonzero or not.

\noindent Case 1: $\epsilon_0=\epsilon_1=0$.

This reduces to $\mathcal{P}_0$ and $\mathcal{P}_1$ (and hence $P_0^{(N)}$ and $P_1^{(N)}$) being known exactly, and Proposition~1 in \cite{Kop:75} has shown that the maximum error rate is bounded above zero if and only if the log-likelihood of $P_0^*$ and $P_1^*$ is bounded from either above or below.

\noindent Case 2: Either $\epsilon_0=0$ or $\epsilon_1=0$, but not both.

In this case, one of the $\mathcal{P}_i$ reduces to the nominal probability distribution $P_i^*$. Without loss of generality, let this be $\mathcal{P}_0$. Define $(Q_0,Q_1)=(P_0^*,Q_1)$ as the LFDs of $\mathcal{P}_0$ and $\mathcal{P}_1$. For $i=0,1$, let the probability density of $P_i^*$ be $p_i^*$. Since $\epsilon_0=0$ and  $\epsilon_1\neq 0$, we have $c''=\infty$ and $c'>0$. Note that for $p_1^*(x)/p_0^*(x)>c'$, we have $l^*(x)= bp_1^*(x)/p_0^*(x)$, where $b=1-\epsilon_1$. Hence, if $p_1^*(x)/p_0^*(x)$ is bounded from above, then for $p_1^*(x)/p_0^*(x)>c'$ we have
\begin{dmath*}
0 < bc' \leq l^*(x)= b \frac{p_1^*(x)}{p_0^*(x)} < \infty.
\end{dmath*}
From Theorem \ref{theorem:LFD_tandem}, we have
\begin{dmath*}
\inf_{\phi^{(N)}}\sup_{(P_0\tn{N},P_1\tn{N})\in\mathcal{P}_0\tn{N}\times\mathcal{P}_1\tn{N}} P_{e,N}(\phi\tn{N}, P_0\tn{N}, P_1\tn{N})
\geq \inf_{\phi^{(N)}}P_{e,N}(\phi\tn{N}, Q_0^{N}, Q_1^{N}),
\end{dmath*}
which is bounded above zero as $N\rightarrow \infty$ since $\log(l^*(x))$ is bounded from both above and below (similar to case 1).
Hence we will assume that the log-likelihood of $P_0^*$ and $P_1^*$ is unbounded from above. Then, the log-likelihood of $Q_0$ and $Q_1$ is bounded from below but not from above as well.

Using the scheme proposed in \cite{PapAth:92}, we can show that we can make the maximum error arbitrarily small as the number of agents tends to infinity. We denote this scheme as $\Phi_\delta$. The decision rules of $\Phi_\delta$ are as follows:

For a given $N^*$ and a threshold $t$,
\begin{dmath*}
U_1=\begin{cases} 0 \textnormal{ for } l^*(Y_1)<t \\ 1 \textnormal{ for } l^*(Y_1)\geq t.\end{cases}
\end{dmath*}
For $1<k<N^*$,
\begin{dmath*}
U_k=\begin{cases} 0 \textnormal { for } l^*(Y_k)<t \textnormal{ and } U_{k-1}=0 \\ 1 \textnormal { otherwise.}\end{cases}
\end{dmath*}
For $k\geq N^*$,
\begin{dmath*}
U_k=U_{k-1}.
\end{dmath*}
Note that $\Phi_\delta$ is a sequence of likelihood ratio tests between $Q_1$ and $Q_0$. It was shown in \cite{PapAth:92} that by choosing a suitable $t$ and $N^*$, we can get an arbitrarily small error rate if $P_1^{(N)}=Q_1^{N}$. To see this, consider a point on the ROC curve of an agent using the likelihood ratio test between $Q_0$ and $Q_1$ with its tangent to the ROC curve having slope $t$. This point is $(Q_0(l^*(Y)\geq t),Q_1(l^*(Y)\geq t))$. As $P_1^*=Q_1$, the initial slope of the ROC curve is $\infty$, and so such a point always exists for any $t$. From the concavity of the ROC curve, we have
\begin{dmath*}
Q_0(l^*(Y)\geq t)<\frac{Q_1(l^*(Y)\geq t)}{t}.
\end{dmath*}
The asymptotic miss detection probability using the decision rules outlined above is thus
\begin{dmath*}
\lim_{N\rightarrow\infty}P_{M,N}(\Phi_\delta, Q_1^N)= Q_1({U_{N^*}=0})
=(1-Q_1({l^*(Y)\geq t)})^{N^*}.
\end{dmath*}
Similarly, the asymptotic false alarm probability is
\begin{dmath*}
\lim_{N\rightarrow\infty}P_{F,N}(\Phi_\delta, Q_0^N)= 1-(1-Q_0({l^*(Y)\geq t}))^{N^*}
<1-(1-\frac{{Q_1(l^*(Y)\geq t})}{t})^{N^*}.
\end{dmath*}
Choose an arbitrary $\delta>0$. To get the asymptotic miss detection probability smaller than $\delta$, we have
\begin{dmath*}
(1-Q_1(l^*(Y)\geq t))^{N^*}<\delta
\end{dmath*}
and so
\begin{dmath}
 N^*>\frac{\log(\delta)}{\log(1-Q_1(l^*(Y)\geq t))}.\label{eq:lower}
\end{dmath}
Similarly, to get the asymptotic false alarm probability smaller than $\delta$, we have
\begin{dmath}
\frac{\log(1-\delta)}{\log(1-\frac{Q_1(l^*(Y)\geq t)}{t})}>N^*.\label{eq:upper}
\end{dmath}
We now make use of the following lemma.
\begin{lemma_A}\label{lemma:jensen}
For any probability distribution $Q_1\in\mathcal{R}$, 
\begin{dmath*}
 \lim_{t\rightarrow \infty} \frac{\log(1-Q_1(l^*(Y)\geq t))}{\log(1-\frac{Q_1(l^*(Y)\geq t)}{t})}=\infty.
\end{dmath*}
\end{lemma_A}
\begin{IEEEproof}
Let $g(x)=\log(1-x)$, a concave function. For $0<x<1$, we have $g(x)<0$.

For any fixed $t>0$, by Jensen's Inequality,
\begin{dmath*}
\frac{1}{t}g(x)+\frac{t-1}{t}g(0)\leq g(\frac{x}{t}). 
\end{dmath*}
As $g(0)=0$, we have
\begin{dmath*}
\frac{g(x)}{g(\frac{x}{t})} \geq t.
\end{dmath*}
Letting $x=Q_1(l^*(Y)\geq t)$,
\begin{dmath*}
\frac{\log(1-Q_1(l^*(Y)\geq t))}{\log(1-\frac{Q_1(l^*(Y)\geq t)}{t})} \geq t.
\end{dmath*}
Hence,
\begin{dmath*}
 \lim_{t\rightarrow \infty} \frac{\log(1-Q_1(l^*(Y)\geq t))}{\log(1-\frac{Q_1(l^*(Y)\geq t)}{t})}=\infty.
\end{dmath*}
\end{IEEEproof}
%
%
%
%
From Lemma \ref{lemma:jensen}, by choosing a large enough $t$, we have
\begin{dmath*}
\frac{\log(1-Q_1(l^*(Y)\geq t))}{\log(1-\frac{Q_1(l^*(Y)\geq t)}{t})} > 2\frac{\log(\delta)}{\log(1-\delta)}
\end{dmath*}
and so
\begin{dmath*}
\frac{\log(1-\delta)}{\log(1-\frac{Q_1(l^*(Y)\geq t)}{t})} > 2\frac{\log(\delta)}{1-\log(Q_1(l^*(Y)\geq t))}.
\end{dmath*}
For a sufficiently large $t$,
\begin{dmath*}
\frac{\log(\delta)}{1-\log(Q_1(l^*(Y)\geq t))}>1.
\end{dmath*}
Hence we can find an integer $N^*$ that lies between the bounds in \eqref{eq:lower} and \eqref{eq:upper}. Then,
\begin{dgroup*}
\dm{\lim_{N\rightarrow\infty}\inf_{\phi^{(N)}} P_N^{\text{DD}}(\phi^{(N)})\leq  \lim_{N\rightarrow\infty}P_N^{\text{DD}}(\Phi_\delta)
=\lim_{N\rightarrow\infty}\sup_{P_0^{(N)}\in\mathcal{P}_0^{(N)}}\pi_0 P_{F,N}(\Phi_\delta,P_0^{(N)})+\pi_1 \sup_{P_1^{(N)}\in\mathcal{P}_1^{(N)}}P_{M,N}(\Phi_\delta,P_1^{(N)})}
\dm[eq:abc]{=\lim_{N\rightarrow\infty}\pi_0 P_{F,N}(\Phi_\delta,Q_0^{(N)})+\pi_1 P_{M,N}(\Phi_\delta,Q_1^{(N)})}
\dm{<\pi_0\delta+\pi_1\delta =\delta,}
\end{dgroup*}
where \eqref{eq:abc} is due to Theorem \ref{theorem:LFD_tandem}. Since $\delta$ was arbitrarily chosen, for this case, we can make the maximum error arbitrarily small as long as one of the following is true:
\begin{enumerate}[1)]
\item $\epsilon_1>\epsilon_0=0$ and the log-likelihood ratio of $P_0^*$ versus $P_1^*$ is unbounded from above,
\item $\epsilon_0>\epsilon_1=0$ and the log-likelihood ratio of $P_0^*$ versus $P_1^*$ is unbounded from below.
\end{enumerate}

\noindent Case 3: $\epsilon_0>0$  and $\epsilon_1>0$.

Here, $c'>0$ and $c''<\infty$ and so the log-likelihood of $Q_1$ and $Q_0$ will be bounded. Then
\begin{dmath*}
\lim_{N\rightarrow\infty}\inf_{\phi^{(N)}}\sup_{(P_0\tn{N},P_1\tn{N})} P_{e,N}(\phi\tn{N}, P_0\tn{N}, P_1\tn{N})
\geq \lim_{N\rightarrow\infty}\inf_{\phi^{(N)}}P_{e,N}(\phi\tn{N}, Q_0^{N}, Q_1^{N}),	
\end{dmath*}
which is known to be bounded above zero as $N\rightarrow \infty$ as the log-likelihood of $Q_1$ and $Q_0$ is bounded from both above and below.

\subsection{Proof of Theorem \ref{theorem:varying}}\label{proof:theorem:varying}
We prove necessity by contradiction. Observe that if such subsequences do not exist, then either $(\epsilon_{0,k})_{k=1}^\infty$ or $(\epsilon_{0,k})_{k=1}^\infty$ is bounded above zero. Without loss of generality, assume that $(\epsilon_{1,k})_{k=1}^\infty$ is bounded above zero. As shown in \cite{Hub:65}, $c'$ as defined in \eqref{eqn:LRLFD} is increasing in $\epsilon_{1,k}$. Therefore, $c'$ is bounded above zero for all $k\geq 1$. This implies that $l^*(Y_{k})$ is bounded above zero as well. Let $\displaystyle\inf_{k,y} l^*(Y_{k} = y)=\delta > 0$. 

Assume now that $\pi_0Q_{F,k}+\pi_1Q_{M,k}\rightarrow 0$ as $k\to 0$. Then both $Q_{F,k}\rightarrow 0$ and $Q_{M,k}\rightarrow 0$. Choose $j$ such that for all $k\geq j$, $Q_{F,k} + Q_{M,k} \leq 1$ and $\pi_0 Q_{F,k} + \delta\pi_1 Q_{M,k} \leq \delta\pi_1$. 
We have from \eqref{varying_rule},
\begin{dmath*}
Q_{F,j+1}
={Q_{F,j} \cdot  Q_{0,{j+1}}\left(l^*(Y_{j+1})\geq\frac{\pi_0Q_{F,j}}{\pi_1(1-Q_{M,j})}\right)}
+{(1-Q_{F,j})\cdot  Q_{0,{j+1}}\left(l^*(Y_{j+1})\geq\frac{\pi_0(1-Q_{F,j})}{\pi_1Q_{M,j}}\right)}
\geq Q_{F,j} \cdot  Q_{0,{j+1}}\left(l^*(Y_{j+1}) \hiderel{\geq} \delta\right)
= Q_{F,j}.
\end{dmath*}
By induction, $Q_{F,k}\geq Q_{F,j}$ for all $k\geq j$, a contradiction. The necessity proof is now complete.

To prove sufficiency, assume that $\epsilon_{0,k_n} \to 0$ and $\epsilon_{1,j_n} \to 0$. We first show a series of lemmas.

\begin{lemma_A}\label{lemma_mean}
Suppose that the LFDs for a pair of uncertainty classes $(\cP_0,\cP_1)$ are $(Q_0,Q_1)$. Then, for any random variable $Y$ with distributions belonging to these uncertainty classes, we have for all $t>0$, 
\begin{dmath*}
Q_1(l^*(Y)\leq t)\leq tQ_0({l^*(Y)\leq t})-\frac{t}{2}Q_0({l^*(Y)\leq t/2}),
\end{dmath*}
and
\begin{dmath*}
Q_0({l^*(Y)\geq t}) \leq \frac{1}{t}Q_1({l^*(Y)\geq t})-\frac{1}{2t}Q_1({l^*(Y)\geq 2t}).
\end{dmath*}
Furthermore, the above two inequalities hold when $l^*(Y)\leq t$ and $l^*(Y)\geq t$ are replaced by $l^*(Y)<t$ and $l^*(Y) >t$, respectively throughout.
\end{lemma_A}
\begin{IEEEproof}
To show the first inequality, we observe that
\begin{dmath*}
Q_1({l^*(Y)\leq t})=tQ_0({l^*(Y)\leq t})-\int_{x=0}^t(t-x)\,dQ_0({l^*(Y)=x})
\leq tQ_0({l^*(Y)\leq t})-\int_{x=0}^{\frac{t}{2}}(t-x)\,dQ_0({l^*(Y)=x})
\leq tQ_0({l^*(Y)\leq t})-\int_{x=0}^{\frac{t}{2}}(t-\frac{t}{2})\,dQ_0({l^*(Y)=x})
=tQ_0({l^*(Y)\leq t})-\frac{t}{2}Q_0({l^*(Y)\leq t/2}).
\end{dmath*}
The second inequality follows by interchanging $Q_0$ and $Q_1$, and replacing $t$ by $1/t$. The proof is now complete.
\end{IEEEproof}

\begin{lemma_A}\label{lemma_decrease}
Suppose that $Q_{F,k}+Q_{M,k}$ is bounded away from zero, and $\epsilon_{0,k}$ or $\epsilon_{1,k}\to 0$ as $k\to\infty$. Then for any integer $n$, there exists some $k'\geq n$ such that \begin{dmath*}
\pi_0Q_{F,{k'}}+\pi_1Q_{M,{k'}}
\leq \pi_0Q_{F,{k'-1}}+\pi_1Q_{M,{k'-1}}-{\frac{\pi_1Q_{M,k'-1}}{2}\cdot Q_{1,{k'}}\left(l^*(Y_{k'})\geq\frac{2\pi_0(1-Q_{F,k'-1})}{\pi_1Q_{M,k'-1}}\right)}
-{\frac{\pi_0Q_{F,k'-1}}{2}\cdot Q_{0,{k'}}\left(l^*(Y_{k'})<\frac{\pi_0Q_{F,k'-1}}{2\pi_1(1-Q_{M,k'-1})}\right).}
\end{dmath*}
\end{lemma_A}
\begin{IEEEproof}
For any $n$, choose $k' \geq n$ such that $Q_{1,k'}\left(l^*(Y_{k'})\geq\frac{2\pi_0(1-Q_{F,{k'-1}})}{\pi_1Q_{M,{k'-1}}}\right)>0$ or 
\\$Q_{0,k'}\left(l^*(Y_{k'})<\frac{\pi_0Q_{F,{k'-1}}}{2\pi_1(1-Q_{M,{k'-1}})}\right)>0$. This is possible as we assume either $Q_{F,k}$ or $Q_{M,k}$ is bounded away from 0, and as $\epsilon_{0,k}$ or $\epsilon_{1,k}\to 0$, the lower or upper bound on $l^*(Y_k)$ converges to 0 or $\infty$, respectively (cf.\ \eqref{eqn:LRLFD}).  Then, we have
\begin{dmath*}
\pi_0Q_{F,{k'}}+\pi_1Q_{M,{k'}}
=\pi_0Q_{F,{k'-1}} \cdot  {Q_{0,{k'}}\left(l^*(Y_{k'})\geq\frac{\pi_0Q_{F,{k'-1}}}{\pi_1(1-Q_{M,{k'-1}})}\right)}
+\pi_0(1-Q_{F,k'-1})\cdot  {Q_{0,{k'}}\left(l^*(Y_{k'})\geq\frac{\pi_0(1-Q_{F,k'-1})}{\pi_1Q_{M,k'-1}}\right)}
+\pi_1Q_{M,{k'-1}}\cdot {Q_{1,{k'}}\left(l^*(Y_{k'})<\frac{\pi_0(1-Q_{F,{k'-1}})}{\pi_1Q_{M,{k'-1}}}\right)}
+\pi_1(1-Q_{M,{k'-1}})\cdot {Q_{1,{k'}}\left(l^*(Y_{k'})<\frac{\pi_0Q_{F,{k'-1}}}{\pi_1(1-Q_{M,{k'-1}})}\right)}
\leq\pi_0Q_{F,k'-1} \cdot  {Q_{0,{k'}}\left(l^*(Y_{k'})\geq\frac{\pi_0Q_{F,k'-1}}{\pi_1(1-Q_{M,k'-1})}\right)}
+\pi_1Q_{M,k'-1} \cdot  {Q_{1,{k'}}\left(l^*(Y_{k'})\geq\frac{\pi_0(1-Q_{F,k'-1})}{\pi_1Q_{M,k'-1}}\right)}
-\frac{\pi_1Q_{M,k'-1}}{2}\cdot {Q_{1,{k'}}\left(l^*(Y_{k'})\geq\frac{2\pi_0(1-Q_{F,k'-1})}{\pi_1Q_{M,k'-1}}\right)}
+\pi_1Q_{M,k'-1}\cdot {Q_{1,{k'}}\left(l^*(Y_{k'})<\frac{\pi_0(1-Q_{F,k'-1})}{\pi_1Q_{M,k'-1}}\right)}
+\pi_0Q_{F,k'-1}\cdot {Q_{0,{k'}}\left(l^*(Y_{k'})<\frac{\pi_0Q_{F,k'-1}}{\pi_1(1-Q_{M,k'-1})}\right)}
-\frac{\pi_0Q_{F,k'-1}}{2}\cdot {Q_{0,{k'}}\left(l^*(Y_{k'})<\frac{\pi_0Q_{F,k'-1}}{2\pi_1(1-Q_{M,k'-1})}\right)}
\end{dmath*}
where the inequality follows from Lemma \ref{lemma_mean}. The lemma is now proved.
\end{IEEEproof}

\begin{lemma_A}\label{lemma:converge}
For $i=0,1$, if $\epsilon_{i,k}\to 0$, then $Q_{i,k}$ converges in distribution to $P_i^*$, where $P_i^*$ is the nominal distribution of the uncertainty class for hypothesis $i$.
\end{lemma_A}
\begin{IEEEproof}
From the definition of the uncertainty classes, we have
\begin{dmath*}
Q_{i,k} \in\left\{{Q \mid  Q =(1-\epsilon_{i,k})P_i^* + \epsilon_{i,k} R}, {R \in \mathcal{R}}\right\}.
\end{dmath*}
For all $x\geq 0$, we have 
\begin{dmath*}
Q_{i,k}({l^*(Y)<x})=(1-\epsilon_{i,k})P_i^*({l^*(Y)<x})+\epsilon_{i,k} R({l^*(Y)<x}),
\end{dmath*}
for some $R\in\mathcal{R}$. Since \begin{dmath*}
(1-\epsilon_{i,k})P_i^*({l^*(Y)<x})\leq (1-\epsilon_{i,k})P_i^*({l^*(Y)<x})+\epsilon_{i,k} R({l^*(Y)<x})\leq (1-\epsilon_{i,k})P_i^*({l^*(Y)<x})+\epsilon_{i,k},
\end{dmath*}
the result follows immediately.
\end{IEEEproof}

We now return to the sufficiency proof of Theorem \ref{theorem:varying}. If $Q_{F,k}+Q_{M,k} \to 0$, the theorem holds trivially. Therefore we assume otherwise. Since $P_k^{\text{SL}}(\phi_k\mid\phi\tn{k-1})$ is bounded and non-increasing, it converges. Suppose that $\pi_0Q_{F,k}+\pi_1Q_{M,k} \to C$, for some $C>0$. Either $\limsup_{k_n}(Q_{F,k_n})>0$ or $\limsup_{k_n}(Q_{M,k_n})>0$. Without loss of generality, let $\limsup_{k_n}(Q_{F,k_n})=C'> 0$. Then there exists a subsequence of agents $(k_{n_\alpha})_{\alpha\geq 1}$ such that $Q_{F,k_{n_\alpha}}\rightarrow C'$. Choose $N$ such that $Q_{F,k_{n_{\alpha'}}}> \frac{C'}{2}$ for all $\alpha'\geq \alpha$. Then, from Lemma \ref{lemma_mean} and Lemma \ref{lemma_decrease}, we have
\begin{dmath*}
P_{k_{n_{\alpha+1}}}^{\text{SL}}({\phi_{k_{n_{\alpha+1}}}\mid\phi\tn{{k_{n_{\alpha+1}}}-1}})
\leq P_{k_{n_{\alpha}}+1}^{\text{SL}}({\phi_{k_{n_{\alpha}}+1}\mid\phi\tn{{k_{n_{\alpha}}}}})
\leq P_{k_{n_{\alpha}}}^{\text{SL}}({\phi_{k_{n_{\alpha}}}\mid\phi\tn{{k_{n_{\alpha}}}-1}})-\frac{\pi_0Q_{F,{k_{n_{\alpha}}}-1}}{2}\cdot {Q_{0,{{k_{n_{\alpha}}}}}\left(l^*(Y_{{k_{n_{\alpha}}}})<\frac{\pi_0Q_{F,{k_{n_{\alpha}}-1}}}{2\pi_1(1-Q_{M,{k_{n_{\alpha}}-1}})}\right)}
\leq P_{k_{n_{\alpha}}}^{\text{SL}}({\phi_{k_{n_{\alpha}}}\mid\phi\tn{{k_{n_{\alpha}}}-1}})-\frac{\pi_0C'}{4}\cdot {Q_{0,{{k_{n_{\alpha}}}}}\left(l^*(Y_{{k_{n_{\alpha}}}})<\frac{C'}{4\pi_1}\right)}.
\end{dmath*}
From Lemma \ref{lemma:converge}, letting $\alpha\to\infty$ on both sides of the above inequality, we obtain a contradiction. The proof of the theorem is now complete.

\subsection{Proof of Theorem \ref{theorem:unknown_social}}\label{proof:unknown_social}
The decision rule of agent 1, $\phi_1$, is of the form
\begin{dmath*}
u_1=\begin{cases}
0\textnormal{ if }l^*(Y_1)<\frac{\pi_0}{\pi_1}
\\1\textnormal{ if }l^*(Y_1)\geq \frac{\pi_0}{\pi_1}.
\end{cases}
\end{dmath*}
Hence, the minimax error probability for the first agent is $\pi_0Q_0(l^*(Y_1)\geq \frac{\pi_0}{\pi_1})+\pi_1Q_1(l^*(Y_1)\geq \frac{\pi_0}{\pi_1})$, which is at least $\inf_\phi P_{\max}^{\text{SL}}(\phi)$. This is because using the trivial rule of ignoring one's own private observation and simply passing on the decision of the previous agent, we get a constant maximum error probability of $\pi_0Q_0(l^*(Y_1)\geq \frac{\pi_0}{\pi_1})+\pi_1Q_1(l^*(Y_1)\geq \frac{\pi_0}{\pi_1})$. Hence, we have
\begin{dmath}
\inf_{\phi}\sup_{k\geq 2}P_k^{\text{SL}}(\phi\mid(\phi_1,\phi))\leq\pi_0Q_0({l^*(Y_1)\geq \frac{\pi_0}{\pi_1}})+\pi_1Q_1({l^*(Y_1)\geq \frac{\pi_0}{\pi_1}}).\label{eq:nonincreasing_error}
\end{dmath}
We next prove the following lemma.
\begin{lemma_A}
Suppose that Assumptions \ref{assumpt:identical} and \ref{assumpt:unknown} hold. There is no loss of optimality if the optimal decision rule $\phi_*$ for all agents $k\geq 2$ is restricted to be a randomized likelihood ratio test between $Q_0$ and $Q_1$. Furthermore, we have
\begin{dmath*}
P_{\max}^{\textrm{SL}}(\phi_*) 
=\sup_{k\geq 2} P_{e,k}((\phi_1,\phi_*), Q_0^k, Q_1^k).
\end{dmath*}
\end{lemma_A}
\begin{IEEEproof}
The proof is similar to that of Lemma \ref{lemma:8}. From the proof of Lemma \ref{lemma:8} it is shown that if $\phi_*$ is not a randomized likelihood ratio test between $Q_0$ and $Q_1$, there is some decision rule $\phi_*'$ which is a randomized likelihood ratio test between $Q_0$ and $Q_1$ such that for any $k\geq 2$, $P_0\tn{k}$, $P_1\tn{k}$,
\begin{dmath*}
P_{e,k}((\phi_1,\phi_*), P_0\tn{k}, P_1\tn{k}) \geq P_{e,k}((\phi_1,\phi_*'), P_0\tn{k}, P_1\tn{k}),
\end{dmath*}
and so there is no loss in optimality in assuming that $\phi_*$ is a likelihood ratio test between $Q_0$ and $Q_1$.

From Theorem \ref{theorem:LFD_tandem}, if $\phi_*$ is a randomized likelihood ratio test between $Q_0$ and $Q_1$, then for any $k\geq 2$, $P_0\tn{k}$, $P_1\tn{k}$, we have
\begin{dmath*}
P_{e,k}((\phi_1,\phi_*), P_0\tn{k}, P_1\tn{k})\leq P_{e,k}((\phi_1,\phi_*), Q_0^k, Q_1^k).
\end{dmath*}
Hence, we obtain
\begin{dmath*}
P_{\max}^{\textrm{SL}}(\phi_*) =\sup_{k\geq 2}\sup_{P_0\in\mathcal{P}_0,P_1\in\mathcal{P}_1}P_{e,k}((\phi_1,\phi_*), P_0\tn{k}, P_1\tn{k})
=\sup_{k\geq 2}P_{e,k}((\phi_1,\phi_*), Q_0^k, Q_1^k),
\end{dmath*}
and the lemma is proved.
\end{IEEEproof}

We now return to the proof of Theorem \ref{theorem:unknown_social}. From \eqref{eq:nonincreasing_error}, we have
\begin{dgroup*}
\dm{P_2^{\text{SL}}(\phi_* \mid \phi_1)\leq \sup_{k\geq 2}P_k^{\text{SL}}({\phi\mid(\phi_1,\phi_*)})}
\dm[eq:2>1]{\leq \pi_0Q_0({l^*(Y_1)\geq \frac{\pi_0}{\pi_1}})+\pi_1Q_1({l^*(Y_1)< \frac{\pi_0}{\pi_1}})}.
\end{dgroup*}
Let
\begin{dgroup*}
\begin{dmath*}
\alpha = P_{F,2}((\phi_1,\phi_*),Q_0^2)-\lim_{k\rightarrow\infty}P_{F,k}((\phi_1,\phi_*),Q_0^k),
\end{dmath*}
\begin{dmath*}
\beta = P_{M,2}((\phi_1,\phi_*),Q_1^2)-\lim_{k\rightarrow\infty}P_{M,k}((\phi_1,\phi_*),Q_1^k).
\end{dmath*}
\end{dgroup*}
and define the function
\begin{dmath*}
g(x)=\lim_{k\to\infty}P_{e,k}((\phi_1,\phi_*), Q_0^k, Q_1^k)+\pi_0\alpha\left(Q_0({\phi_*(Y_1,1)=1})-Q_0({\phi_*(Y_1,0)=1})\right)^{x-1}
+\pi_1\beta\left({Q_1(\phi_*(Y_1,0)=0})-Q_1(\phi_*({Y_1,1)=0})\right)^{x-1}.
\end{dmath*}
From the recurrence relation in Lemma \ref{lemma:recurrence}, we see that for integer values of $x$, this function gives $P_{e,x}((\phi_1,\phi_*),Q_0^x,Q_1^x)$. Differentiating with respect to $x$, we observe that $g(x)$ has at most one stationary point for $x>0$. From \eqref{eq:2>1}, we have $g(2)\geq g(1)$. Hence, the stationary point, if any exists, must be a minimum. This means that for $x\geq 2$, the maximum value of $g(x)$ must be either at $x=2$ or at its asymptotic limit. Thus, the theorem is now proven.

\bibliographystyle{IEEEtran}
\bibliography{IEEEabrv,biblio,social_sensing}

\end{document}